\documentclass[12pt]{article}
\usepackage[a4paper,left=30mm,right=20mm,top=20mm,bottom=20mm]{geometry}

\newcommand{\bea}{\begin{eqnarray}}
\newcommand{\eea}{\end{eqnarray}}
\newcommand{\be}{\begin{equation}}
\newcommand{\ee}{\end{equation}}

\usepackage{bm}
\usepackage{epsfig}
\usepackage{amssymb}
\usepackage{amsmath}
\usepackage{calrsfs}
\usepackage{multirow}
\usepackage{rotating}
\usepackage{xcolor}

\begin{document}

\setlength{\arraycolsep}{0.5mm}

\title{$\alpha_s$ from hadron multiplicities: an updated analysis
%SUSY-like relation in evolution of gluon and quark jet multiplicities
}
  \author{A.V.~Kotikov$^{1}$, B.G.~Shaikhatdenov$^{1}$, N.D.~Godlevsky$^{2}$, I.A.~Kotikov$^{2}$\\
%\thanks{Permanent address: 
    $^{1}$ Joint Institute for Nuclear Research, 141980 Dubna, Russia\\
    $^{2}$ Moscow State University, Faculty of Physics, 119991, Moscow, Russia
}

\maketitle

\begin{abstract}
Exploring 
%  We use the  relationship \cite{Kniehl:2017fix} between
the anomalous
dimensions, resummed through next-to-next-to-leading-logarithmic order,
we study
%Dokshitzer-Gribov-Lipatov-Altarelli-Parisi (DGLAP) evolution equations for
the first Mellin moments $D_{q,g}(\mu^2)$ of the quark and gluon fragmentation
functions, which correspond to the average hadron multiplicities in jets
initiated by quarks and gluons, respectively.
By fitting the world data on $D_{q,g}(\mu^2)$ for charged hadrons measured in $e^+e^-$
annihilation, we find a good agreement with the data and obtain the following
value for the strong force coupling
$\alpha_s^{(5)}(M_Z)=0.1177\genfrac{}{}{0pt}{}{+0.0027}{-0.0025}$. 

\end{abstract}

%\pacs{12.38.Cy,12.39.St,13.66.Bc,13.87.Fh}

%\maketitle
\section{Introduction}
%In the parton model of QCD \cite{Bjorken:1969ja}, 
The analysis of inclusive
production of single hadrons in the framework of the parton model of QCD~\cite{Bjorken:1969ja}
involves the notion of fragmentation functions %(FFs)
$D_a(x,\mu^2)$  (hereafter $(a=q,g)$, $\mu$ is a factorization scale).
Their $\mu^2$ dependences are governed by the timelike
Dokshitzer-Gribov-Lipatov-Altarelli-Parisi (DGLAP) evolution 
equations~\cite{Gribov:1972ri,Dokshitzer:1977sg}.

The DGLAP equations can be analytically solved in Mellin space, where
$D_a(N,\mu^2)=\int dx\,x^{N-1}D_a(x,\mu^2)$ 
%with 
$(N=1,2,\ldots)$, 
%and similarly for $P_{ba}(x)$, 
thanks to the convolutions getting converted into products
\begin{equation}
\frac{\mu^2d}{d\mu^2}
\left(\begin{array}{l} D_s(N,\mu^2) \\ D_g(N,\mu^2) \end{array}\right)
=\left(\begin{array}{ll} P_{qq}(N) & P_{gq}(N) \\
P_{qg}(N) & P_{gg}(N) \end{array}\right)
\left(\begin{array}{l} D_s(N,\mu^2) \\ D_g(N,\mu^2) \end{array}\right),
\label{apR}
\end{equation}
where $P_{ab}(N)$ 
%(hereafter 
$(a,b=q,g)$
%) 
are anomalous dimensions and $D_q\sim D_s$, where $D_s$
%$D_s=(1/2n_f)\sum_{q=1}^{n_f}(D_q+D_{\bar{q}})$, with $n_f$ being the number of active quark flavors,
is a quark singlet component (note that a quark non-singlet component
is irrelevant for the present study).

In this paper we are interested in the first Mellin moment $D_a(\mu^2)\equiv D_a(1,\mu^2)$
because, up to corrections of orders beyond our
consideration here, it corresponds to the average hadron multiplicity
$\langle n_h\rangle_a$ of jets initiated by parton $a$.
There already is a database collected by various experiments on $\langle n_h\rangle_q$,
$\langle n_h\rangle_g$, and their ratio
$r=\langle n_h\rangle_g/\langle n_h\rangle_q$ for charged hadrons $h$ taken in
$e^+e^-$ annihilation at various center-of-mass energies $\sqrt{s}$, ranging
from 10 to 209~GeV  (see~\cite{Bolzoni:2013rsa} and references therein).
The study of $D_a$ is known to be a long lasting challenge; the LO value of $r$,
$C^{-1}=C_A/C_F$ with color factors $C_F=4/3$ and $C_A=3$, was found nearly five
decades ago~\cite{Brodsky:1976mg}.

The description of the $\mu^2$ dependences of $D_a$ at fixed orders in
perturbation theory is spoiled by the observation that $P_{ba}\equiv P_{ba}(1)$ are
ill defined and require resummation, which was performed for the leading
logarithms %(LL)
\cite{Mueller:1981ex}, next-to-leading logarithms %(NLL)
\cite{Vogt:2011jv}, and next-to-next-to-leading logarithms (NNLL)
\cite{Kom:2012hd}.

In Ref. \cite{Kniehl:2017fix} (see also  \cite{Kniehl:2017oat}) % and \cite{Kotikov:2020ukv}),
%this Letter, 
we found a relationship between the NNLL-resummed expressions
for $P_{ba}$, which had previously slipped attention.
%which had gone unnoticed so far.
Its existence in QCD is quite remarkable and interesting in its own,
because a similar relationship is known to hold in supersymmetric %(SUSY)
QCD, where $C=1$
\cite{Dokshitzer:1977sg,Kom:2012hd}.
Owing to this new relationship, the DGLAP equations can be solved exactly,
which greatly consolidates the theoretical foundation for the determination of
$\alpha_s$ and thus reduces its theoretical uncertainty.
Moreover, the relation leads to probabilistic properties (see \cite{Kotikov:2020ukv})
 of the properly rescaled parton jet multiplicities obtained from standard ones 
by extracting the quark and gluon ``color charges'' $C_F$ and $C_A$, respectively.

\section{Basic formulas}

%{1.}
Our starting point is Eq.~(\ref{apR}) for $N=1$ with NNLL resummation.
We have \cite{Kom:2012hd}
\begin{eqnarray}
P_{aa}&=&\gamma_0(\delta_{ag} + K_{a}^{(1)} \gamma_0 
+ K_{a}^{(2)}  \gamma_0^2) ,~~
P_{gq}=
%&=& 
C (P_{gg} +A),~~
P_{qg} =
%&=& 
C^{-1} (P_{qq} +A) ,
%+ \mathcal{O}(\gamma_0^4),
\label{NNLL}
\end{eqnarray}
with $\mathcal{O}(\gamma_0^3)$ accuracy,
where $\gamma_0=\sqrt{2C_Aa_s}$, 
%with 
$a_s=\alpha_s/(4\pi)$ is
%being 
the coupling,
$\delta_{ab}$ is the Kronecker symbol, and
\begin{eqnarray}
K_{q}^{(1)} &=& \frac{2}{3} C\varphi,\quad K_{g}^{(1)} =
%&=& 
-\frac{1}{12}[11 +2\varphi (1+6C)],\quad
K_{q}^{(2)} = -\frac{1}{6} C\varphi [17-2\varphi(1-2C)],
\nonumber \\
%K_{g}^{(1)} &=& -\frac{1}{12}[11 +2\varphi (1+6C)],\quad 
K_{g}^{(2)} &=& \frac{1193}{288} -2\zeta_2
- \frac{5\varphi}{72}(7-38C)+\frac{\varphi^2}{72}(1-2C)(1-18C),~
%\nonumber \\
A =
%&=& 
K_{q}^{(1)}\gamma_0^2,~
\varphi=\frac{n_f}{C_A},
\label{nllfirstA}
\end{eqnarray}
where $n_f$ is the number of active quark flavors.

From Eq.~(\ref{NNLL}) it is seen that there is a relationship \cite{Kniehl:2017fix}
\begin{equation}
C^{-1}P_{gq}-P_{gg}=CP_{qg}-P_{qq},
\label{Basic}
\end{equation}
which is independent of $n_f$.

%We now solve
Eq.~(\ref{apR}) can be solved exactly at $N=1$ by exploiting~(\ref{Basic}). Details of the study
%procedure to obtain the solution can
can be found in \cite{Kniehl:2017fix}. Here we restrict ourselves to only the final results.

%Now we can study
The quark and gluon average multiplicities contain (with proper diagonalization) the two components
\be
%D^{\rm pt}_a
D_a=D_a^++D_a^-,~~
D_g^{\pm}=r_{\pm}\,D^{\pm}_q\,,
\label{Da_pm1}
\ee
where
\begin{equation}
\frac{D_q^-(\mu^2)}{D_q^-(\mu_0^2)} =
\exp\!{\left[\int_{\mu_0^2}^{\mu^2}\!\!\!\frac{d\bar{\mu}^2}{\bar{\mu}^2}
\!\left(\frac{C\varphi\beta_0}{3C_A}\gamma_0^3-A\!\right)\!\right]}
= \frac{T_-(\gamma_0(\mu^2))}{T_-(\gamma_0(\mu_0^2))}, ~~
T_-(\gamma_0)=
\gamma_0^{d_-}\exp{\left(-\frac{4}{3}C\varphi \gamma_0\right)}%,
\label{gensol.-}
\end{equation}
and
\begin{eqnarray}
&&\frac{D_q^+(\mu^2)}{\tilde{D}_q^+(\mu_0^2)}=
%&=&
\exp{\left[\int_{\mu_0^2}^{\mu^2}\frac{d\bar{\mu}^2}{\bar{\mu}^2}
\gamma_0\Bigl(1+ (2K_{q}^{(1)}+K_{g}^{(1)})
%K_+^{(1)}
\gamma_0+(K_{q}^{(2)}+K_{g}^{(2)})
%K_+^{(2)}
\gamma_0^2\Bigr)\right]}
%\nonumber\\&=&
=\frac{T_+(\gamma_0(\mu^2))}{T_+(\gamma_0(\mu_0^2))}, \nonumber \\
&&T_+(\gamma_0)
=\gamma_0^{d_+}\exp{\left[\frac{4C_A}{\beta_0\gamma_0}
-\frac{4C_A}{\beta_0} \left(K_{q}^{(2)}+K_{g}^{(2)}
%K_+^{(2)}
-b_1\right)\gamma_0\right]},~~\beta_1=\frac{2C_A^2}{3}[17-\varphi(5+3C)],
\label{gensol.+}
\end{eqnarray}
with $d_-=8C_AC\varphi/(3\beta_0)$,
$d_+=-4C_A(2K_{q}^{(1)}+K_{g}^{(1)})/
%K_+^{(1)}/
\beta_0$ and $b_1=\beta_1/(2C_A \beta_0)$.

Strictly speaking, the ``$+$'' component also contains an inhomogeneous term, whose contribution is negligible (see \cite{Kniehl:2017fix})
and we omit it in the present analysis.

The results for $r_{\pm}$ have the following form~\footnote{The terms 
of $\sim \gamma_0^2$ and  $\sim \gamma_0^3$ in $r_+$ are only known when $n_f=5$ 
(see Ref.\cite{Capella:1999ms}).
  This leads to the usage of $n_f=5$ in the fits of experimental data done below.}:
%\textcolor{blue}{(for $n_f=5$)}and
\bea
&&r_+=2.250-4.505\,\gamma_0^2-0.586\,
\gamma_0^3 +\mathcal{O}(\gamma_0^4)~~\mbox{(for $n_f=5$)};
\label{rp}\\
&&r_{-}=\frac{1-\alpha}{\epsilon}
=-\frac{4}{3}\varphi\gamma_0
+\frac{\varphi}{18}[29-2\varphi(5-2C)]\gamma_0^2
+\mathcal{O}(\gamma_0^3).
\label{rm}
\eea
Following \cite{Kniehl:2017fix},
we denote the approximation of using Eqs.~(\ref{rp}) and (\ref{rm})
%on top of Eqs.~(\ref{eq:rpm}) and (\ref{eq:r})
as NNNLO${}_\mathrm{approx}$+NNLL.

Power-like corrections were found to be indispensable for a realistic
description of the experimental data on
$\langle n_h\rangle_q$ and $\langle n_h\rangle_g$ %, and $r$
\cite{Capella:1999ms,Dokshitzer:1992iy}.
%Following Refs.~\cite{Dokshitzer:1992iy}
Since there was previously a problem \cite{Capella:1999ms} with describing the experimental data on quark multiplicity, we modify the quark part
\be
D_q\to %D^{\rm pt}_s
D_q + \frac{A_q(\mu)}{\mu+\mu_q}
\label{Da1}
\ee
with arbitrary parameters $\mu_q$ and $A_q(\mu)$, which should be obtained while fitting experimental data.
  The magnitude $A_q(\mu)$ will be used in two
  %three
  different forms (see \cite{Dokshitzer:1992iy}):
\be
A_q(\mu)=A_q(\mu_0)
~~\mbox{and}
~~A_q(\mu)=A_q(\mu_0)
%A(\mu_0)
\frac{\gamma^2_0(\mu)}{\gamma^2_0(\mu_0)}
\,.
\label{DaN}
\ee

The power-like form $(\mu+\mu_q)^{-1}$ in the r.h.s of (\ref{Da1}) is taken in close analogy with the ``massive'' form of the high twist corrections
\cite{Teryaev:2013qba}, which was successfully used in recent analyses: to study the polarized Bjorken and Gross-Llewellyn Smith sum rules
(see \cite{Gabdrakhmanov:2023rjt} and  \cite{Gabdrakhmanov:2025vxw}, respectively) and for the pion-photon transition form factor \cite{Zemlyakov:2026rhe}.

When the values of $\mu$ are high enough, 
%In the case where
the form $(\mu+\mu_q)^{-1}$ 
%massive twist-four term in (\ref{TFF_0_NNLO_A_HT})
can be expanded in an inverse series in $\mu^2$.
For smaller values of $\mu^2$, this expansion becomes inapplicable, but the %massive
form shown in (\ref{Da1})  can be used without problems. 

We note that the power-like corrections (\ref{Da1}) are taken in a different way 
as compared to those in \cite{Kniehl:2017fix}, where they were added to
the ``$+$''-component, which boils down to basic contributions mostly at high $\mu^2$ values. 
At low $\mu^2$ values, where the power-like corrections
give valuable contributions, both ``$+$'' and ``$-$''components are important.
%Moreover, namely the quark multiplicities had strong disagreement with the corresponding experimental data.
Indeed, while considering both components in our previous studies \cite{Bolzoni:2013rsa,Bolzoni:2012ii},
the results of comparison between experimental data and theoretical predictions for the quark multiplicities 
was strongly improved at rather high $\mu^2$ values. % with the ``$-$''-components taken in 
consideration~\footnote{It is in full agreement with similar studies of deep inelastic structure functions done 
in Ref. \cite{Kotikov:1998qt}.}.
In order to ensure a good agreement with experimental data for low $\mu^2$ values as well
the power-like corrections are necessary.

Motivated by Refs.~\cite{Brodsky:1976mg,Aversa:1990uv,Aguilar:2014tka} 
and the recent analysis \cite{Subotic:2026oio}, we choose $\mu^2=R^2Q^2+m_\mathrm{eff}^2$, 
where $R$ is a jet radius,
$Q=\sqrt{s}$, and $m_\mathrm{eff}$ is an effective gluon mass.
We adopt $R=0.4$ as a typical value (see recent paper \cite{Karkkainen:2026yak} and references therein)
%from Ref.~\cite{Aversa:1990uv}
and take $m_\mathrm{eff}=m_0=1$~GeV 
and $m_\mathrm{eff}^2=\overline{m}^2_\mathrm{eff}(\mu)=m_0^2/[1+(\mu^2/\Lambda_m^2)^\gamma]$ with %$m=0.375$~GeV,
$\Lambda_m=0.67$~GeV, and $\gamma=1.96$ from Ref.~ \cite{Aguilar:2014tka,Subotic:2026oio}.

%\footnotesize
%\small
%\footnotesize
\scriptsize
\begin{table}[t]%[!htb]%[t]
\centering 
  \begin{center}
%\scriptsize
\footnotesize
    \begin{tabular}{|c|c|c|c|c|c|c|}
\hline
%\footnotesize
%{\small 
$M$ & $D_g(\mu_0^2)$ & $D_q(\mu_0^2)$ & $A_q$ [GeV]& $\mu_q$ [GeV] & $\alpha_s(M_z)$ & $\chi^2/(d.o.f)$ \\
%/({\rm d.o.f.})$   \\
%& (for PT) & (for PT) &  (for PT)\\
\hline 
\footnotesize
$m_0$ & 23.75$\pm$0.22 & 19.61$\pm$0.44 & -108.04$\pm$19.65 & 13.25$\pm$2.05 & $0.1194\genfrac{}{}{0pt}{}{+0.0028}{-0.0026}$ %-0.0026+0.0028
& 1.10 \\
          & (23.75$\pm$0.10) & (20.19$\pm$0.73) & (-316.99$\pm$116.84) & (78.53$\pm$20.86) & ($0.1167\genfrac{}{}{0pt}{}{+0.0037}{-0.0034}$)& (1.47) \\
\hline
$\overline{m}_{\rm eff}(\mu)$ & 23.73$\pm$0.10 & 19.99$\pm$0.49 & -131.83$\pm$24.31 & 16.13$\pm$2.38 & $0.1159\genfrac{}{}{0pt}{}{+0.0026}{-0.0023}$ %-0.0023+0.0026
& 1.07 \\
          & (23.74$\pm$0.10) & (19.89$\pm$0.64) & (-277.01$\pm$92.71) & (74.14$\pm$17.57) & ($0.1162\genfrac{}{}{0pt}{}{+0.0035}{-0.0031}$) & (1.78) \\
%}
\hline
\end{tabular}
\end{center}
\vspace{-0.3cm}
\caption{The fit of initial conditions of $\mu^2$-dependent quark and gluon multiplicities following (\ref{Da_pm1})-
  %, (\ref{gensol.+a}), (\ref{Da}) and
  (\ref{Da1}).
  %parameters for the $\mu^2$-independent $\mu_{\rm{A},4}$ with $b_n(Q_0^2)$ given in (\ref{LA})-(\ref{SP}).
  The results for $A_q(\mu) \sim \gamma^2_0(\mu)$
  are given in brackets.%, wavy brackets, and square brackets, respectively.
  }
\label{Tab:Ind}
\end{table}
\normalsize

\section{Results}

Let us present the results of global fits to the available data taken in measurements
of $\langle n_h\rangle_q$ and $\langle n_h\rangle_g$ for charged hadrons $h$ in
$e^+e^-$ annihilation, which were compiled in
Ref.~\cite{Bolzoni:2013rsa}.
As in Ref.~\cite{Bolzoni:2013rsa,Kniehl:2017fix}, we choose the reference
scale to be $Q_0=50$~GeV, which approximately corresponds to the geometric mean of
the smallest and largest of the occurring $\sqrt{s}$ values, and take $n_f=5$
throughout our analysis.

\begin{figure}
  \centering
\includegraphics[scale=0.8,clip]{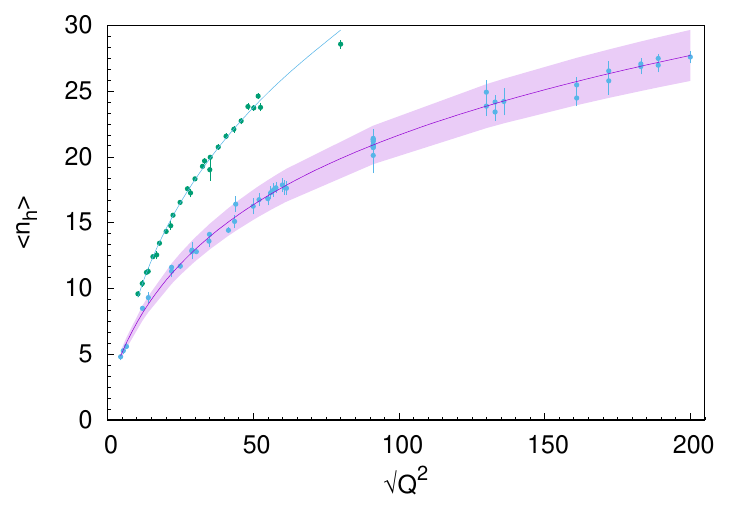}
\vskip -0.5cm
\caption{Comparison between experimental data of $\langle n_h(Q^2)\rangle_q$
(the lower curve) and $\langle n_h(Q^2)\rangle_g$ (the upper curve) and their
$\mathrm{NNNLO}_\mathrm{approx}+\mathrm{NNLL}$ fits.}
\label{fig:n}
\vskip -0.3cm
\end{figure}

As can be seen in Fig.~1 \footnote{In Figs.~1 and 2, an effective gluon mass $m_\mathrm{eff}^2=\overline{m}^2_\mathrm{eff}(\mu)$, 
and $A_q$ parameter is taken to be constant.}
and Table~1
%\ref{fig:n},
our
%$\mathrm{NNNLO}_\mathrm{approx}+\mathrm{NNLL}$
fits yield an excellent
description of the experimental data it includes, with a $\chi^2$ per degree
of freedom of $\chi_\mathrm{dof}^2 \leq 1.1$ in the case of the $\mu$-independent 
magnitude $A_q$ of power-like corrections.
%\textcolor{blue}{
Table 1 shows that gluon multiplicities are described with significantly higher accuracy than quark multiplicities;
  the uncertainties for the latter are shown in Fig. 1 by the shaded band.%}  

Taking the average values of the parameters from fits
%these fits
with different choices of a gluon mass $m(\mu)$, we have
%The fit parameters are determined to be
$\langle n_h(Q_0^2)\rangle_q\equiv D_q(\mu_0^2)+A_q(\mu_0)/(\mu_0+\mu_q)=16.34\pm 1.13$,
$\langle n_h(Q_0^2)\rangle_g\equiv D_g(\mu_0^2)=23.64\pm 0.10$, $A_q=-119.94\pm 21.98$, $\mu_q=14.69\pm 2.22$
(with $\mu_0^2=R^2Q_0^2+m_\mathrm{eff}^2$)
%$K_\mathrm{cr}=7.09\genfrac{}{}{0pt}{}{+1.71}{-1.21}$,
and
\begin{equation}
\alpha_s^{(5)}(M_Z^2)=0.1177\genfrac{}{}{0pt}{}{+0.0027}{-0.0025},
\label{eq:as}
\end{equation}
which is in a good agreement with the present world average,
$\alpha_s(M_Z^2)=0.1180\pm0.0009$ \cite{ParticleDataGroup:2026mpi}  (for a review see
\cite{dEnterria:2022hzv}).
%Our fit results turn out to be very insensitive to the precise choice of $Q_0$ (???).
Also, the average multiplicities $\langle n_h(Q_0^2)\rangle_q$ and $\langle n_h(Q_0^2)\rangle_g$ 
are highly compatible with the corresponding results of previous 
studies \cite{Bolzoni:2013rsa,Kniehl:2017fix,Bolzoni:2012ii,Perez-Ramos:2013eba}.

\begin{figure}
\centering
\includegraphics[scale=0.8,clip]{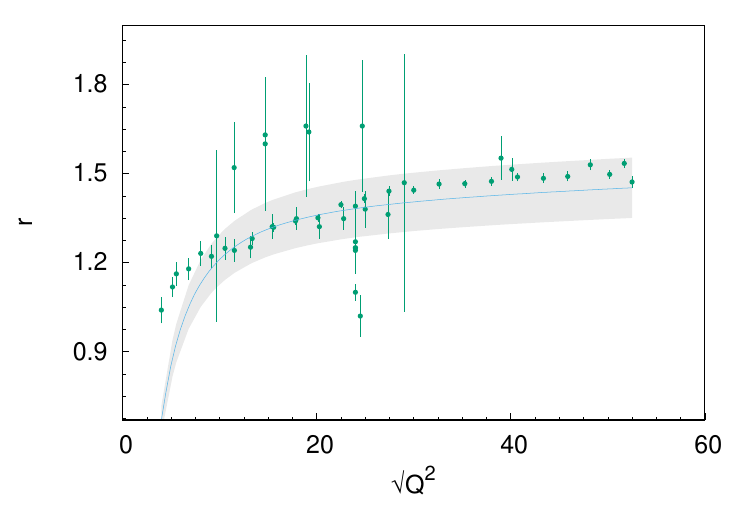}
\vskip -0.5cm
\caption{Comparison between our $\mathrm{NNNLO}_\mathrm{approx}+\mathrm{NNLL}$
prediction for $r(Q^2)$ and experimental data excluded from the fit.}
\label{fig:r}
%\vskip -0.7cm
\vskip -0.2cm
\end{figure}

In Fig.2,
%\ref{fig:r},
we compare our
$\mathrm{NNNLO}_\mathrm{approx}+\mathrm{NNLL}$ prediction for the ratio $r(Q^2)$:
\be
r(Q^2)=\frac{\langle n_h(Q^2)\rangle_g}{\langle n_h(Q^2)\rangle_q}=
\frac{D_g(\mu^2)}{D_q(\mu^2)+A_q(\mu)/(\mu+\mu_q)}\,,%=\frac{r_++r_{-}D^{-}_q(\mu^2)/D^{+}_q(\mu^2)}{1+D^{-}_q(\mu^2)/D^{+}_q(\mu^2)}\,,
\label{r}
\ee
with the experimental data compiled in Ref.~\cite{Bolzoni:2013rsa}, which did not enter
our fits. 
%\textcolor{blue}{
The uncertainties for $r(Q^2)$ are primarily due to uncertainties in the values of quark multiplicities
  and are represented in Fig. 2 by a shaded band.%} 
With the exception of the region of small $Q^2$ values, the agreement can be considered quite satisfactory, which confirms the validity of our analysis.
At small $Q^2$ values, the experimental data for the multiplicities themselves and for their ratio do not appear to be consistent.

This research was supported in part by the Russian
Science Foundation grant No. 25-22-0057.

\end{document}